\documentclass{article}

\usepackage{amsmath}
\usepackage{amssymb}
\usepackage{geometry}
\usepackage{mathtools}
\usepackage{verbatim}
\usepackage[percent]{overpic}
\usepackage{rotate}
\usepackage{caption}
\usepackage{hyperref}

\usepackage{tikz}

\newcommand{\foog}[1]{%
\begin{tikzpicture}[#1]%
\draw[line cap=round] (0.0,0.0ex) -- (0.7ex,0.7ex);%
\draw[line cap=round] (0.7ex,0.0ex) -- (0.0ex,0.7ex);%
\end{tikzpicture}%
}

\newcommand{\foop}[1]{%
\begin{tikzpicture}[#1]%
\draw[line cap=round] (0.0,-0.1ex) -- (0.0ex,0.8ex);%
\draw[line cap=round] (0.25ex,-0.1ex) -- (0.25ex,0.8ex);%
\end{tikzpicture}%
}

\newcommand{\foo}[1]{%
\begin{tikzpicture}[#1]%
\draw[line cap=round] (0,0) -- (0.7ex,0.7ex);%
\end{tikzpicture}%
}

\newcommand{\foos}[1]{%
\begin{tikzpicture}[#1]%
\draw[line cap=round] (0,0.7ex) -- (0.7ex,0);%
\end{tikzpicture}%
}

\begin{document}

\title{Voronoi chains, blocks, and clusters \\in perturbed square lattices}  
\author{Emanuel A. Lazar\thanks{Department of Mathematics, Bar-Ilan University, Ramat Gan 5290002, Israel}\hspace{2mm}\thanks{Email: \href{mailto:mLazar@math.biu.ac.il}{mLazar@math.biu.ac.il}} \and Amir Shoan\footnotemark[1]}  
\date{\today}
%Random Geometry,Exact results,Random/ordered microstructure

\maketitle

\begin{abstract}
\begin{center}
\includegraphics[trim={14mm 14mm 14mm 14mm},clip,width=0.99\linewidth]{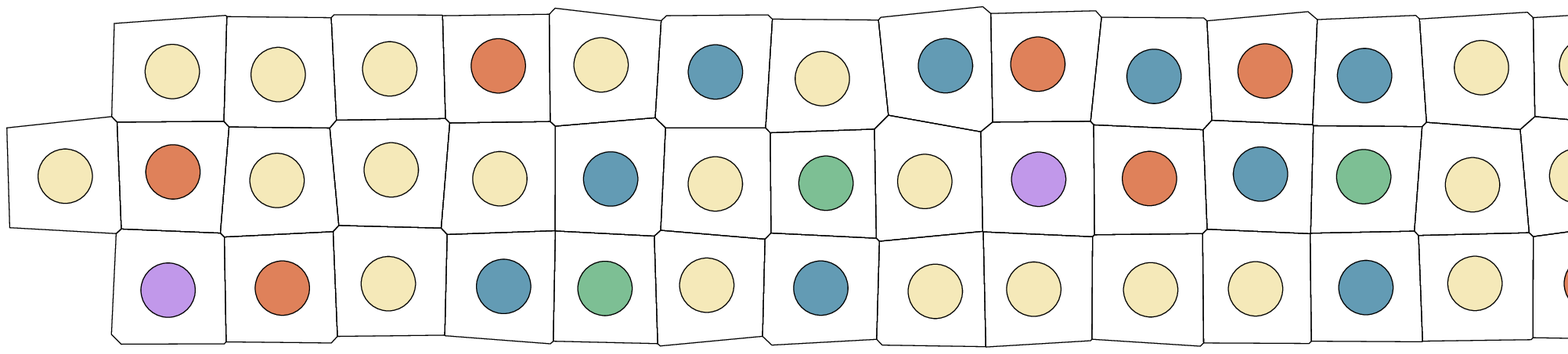}
\end{center}
Perturbed lattices provide simple models for studying many physical systems.  In this paper we study the distribution of Voronoi chains, blocks, and clusters with prescribed combinatorial features in the perturbed square lattice, generalizing earlier work.  In particular, we obtain analytic results for the presence of hexagonally-ordered regions within a square-ordered phase.  Connections to high-temperature crystals and to non-equilibrium phase transitions are considered.  In an appendix, we briefly consider the site-percolation threshold for this system.
\begin{center}
\includegraphics[trim={14mm 14mm 14mm 14mm},clip,width=0.99\linewidth]{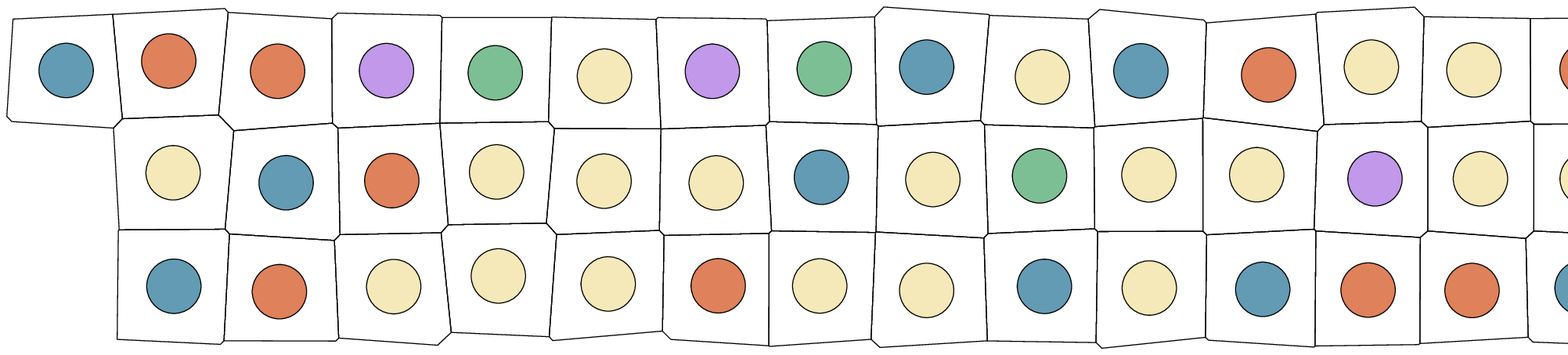}
\end{center}
\end{abstract}

\section{Introduction} 
\subsection{Motivation}

Many physical systems can be modeled as perturbed lattices, and so understanding the manner in which particles are arranged in such systems provides a natural path towards investigating a wide range of physical systems and mechanisms.  Statistical topology studies statistical properties of topological descriptions of systems, and thus provides a natural setting in which to study both perfect and imperfect crystals, much in the way that statistical mechanics focuses on geometric quantities such as velocities and momenta of large sets of particles.  Statistical topology has been previously used to study glasses \cite{rivier1983statistical}, polymers \cite{orlandini2007statistical}, polycrystalline metals \cite{keller2014comparative}, radial and cellular networks \cite{seong2012statistical, mason2012statistical}, ideal gases \cite{lazar2013statistical}, and supercritical fluids \cite{yoon2018topological,yoon2019topological,yoon2019topological2}.  

This paper builds on earlier work that computed the distribution of Voronoi topologies in perturbed two-dimensional crystals \cite{leipold2016statistical}.  The authors illustrated how simple probabilistic analysis can be used to determine the distribution of Voronoi cells, as classified by their number of edges, in such systems.  In particular, they showed that the Voronoi cells in perturbations of two-dimensional Bravais lattices have between 4 and 8 edges, with 6 being the most common in each.  In this paper we consider localized regions of Voronoi cells with prescribed combinatorial features in perturbed square lattices; we call such arrays {\it chains}, {\it blocks}, and {\it clusters}, as detailed below.  With what probabilities will a set of adjacent Voronoi cells, for example, all be hexagonal after a random perturbation of the lattice?

Of course, it is well-known that the square lattice is unstable under most interatomic potentials, and that the ground states of most two-dimensional physical systems are triangular \cite{lozovik2019spontaneous}.  Nonetheless, it has also been shown that certain pair potentials produce classical ground states that are indeed square \cite{lozovik2019spontaneous,marcotte2011optimized}, and that such potentials are possibly realized in some colloidal and polymeric systems \cite{el2008ground}.  The work presented here sheds light on the latent structure in these systems.  

The analysis presented here complements a rich and valuable literature devoted to geometric properties of perturbed lattices and their Voronoi cells.  For example, how are areas, perimeters, volumes, and other geometric features distributed?  Geometric studies of perturbed lattices in two \cite{liao1999voronoi,lucarini2008symmetry} and three \cite{liao2001description,troadec1998statistics,lucarini2009three} dimensions provide an important complementary viewpoint from which to study many important physical and chemical systems.

\subsection{Perturbed lattices}

We model finite-temperature crystals as square lattices that are randomly perturbed in the following sense.  Cartesian coordinates of each point in $\mathbb{Z}^2$ are displaced by independently and identically distributed random variables sampled from a normal distribution $\mathcal{N}(0,\sigma^2)$ centered at the origin and with variance $0 < \sigma^2 \ll 1$.  The resulting {\it perturbed lattice}
\begin{equation}
\Lambda = \left\{ z + \xi_z  | \; z \in \mathbb{Z}^2, \hspace{1mm} \xi_z \sim \mathcal{N}^2(0,\sigma^2) \right \},
\end{equation}
in which the random variables $\xi_z$ are identically and independently sampled, can be considered an approximation of a two-dimensional crystal whose atoms are displaced from lattice positions by thermal noise \cite{holroyd2013insertion, peres2014rigidity}.  Because physical models are the inspiration for this study, we will oftentimes refer to points in the perturbed lattice as atoms.

\subsection{Voronoi tessellations}
\label{Voronoi tessellations}

\begin{figure}
\begin{center}$
\begin{tabular}{cc} 
\fbox{\includegraphics[trim={14mm 14mm 14mm 14mm},clip,width=0.27\linewidth]{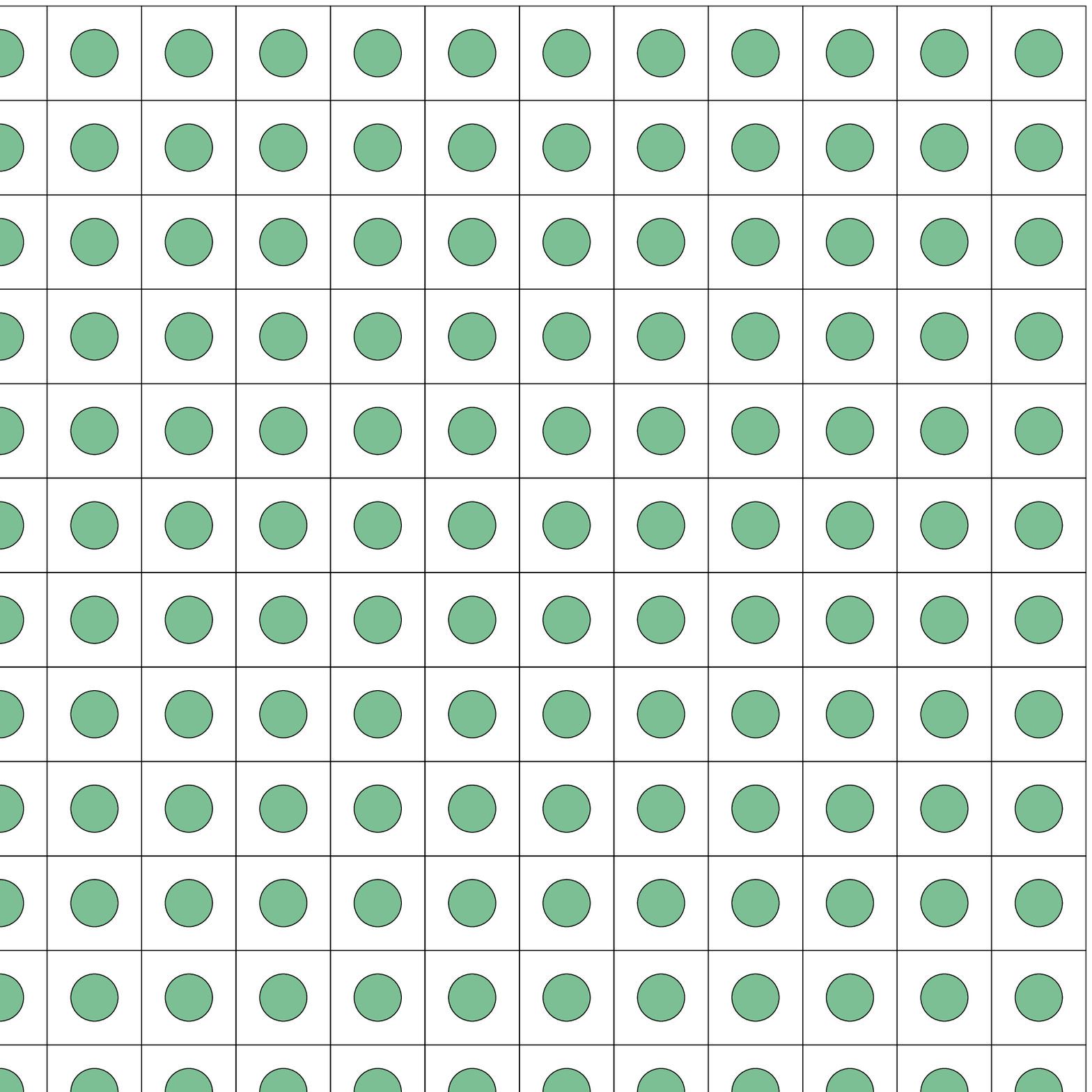}} &
\fbox{\includegraphics[trim={14mm 14mm 14mm 14mm},clip,width=0.27\linewidth]{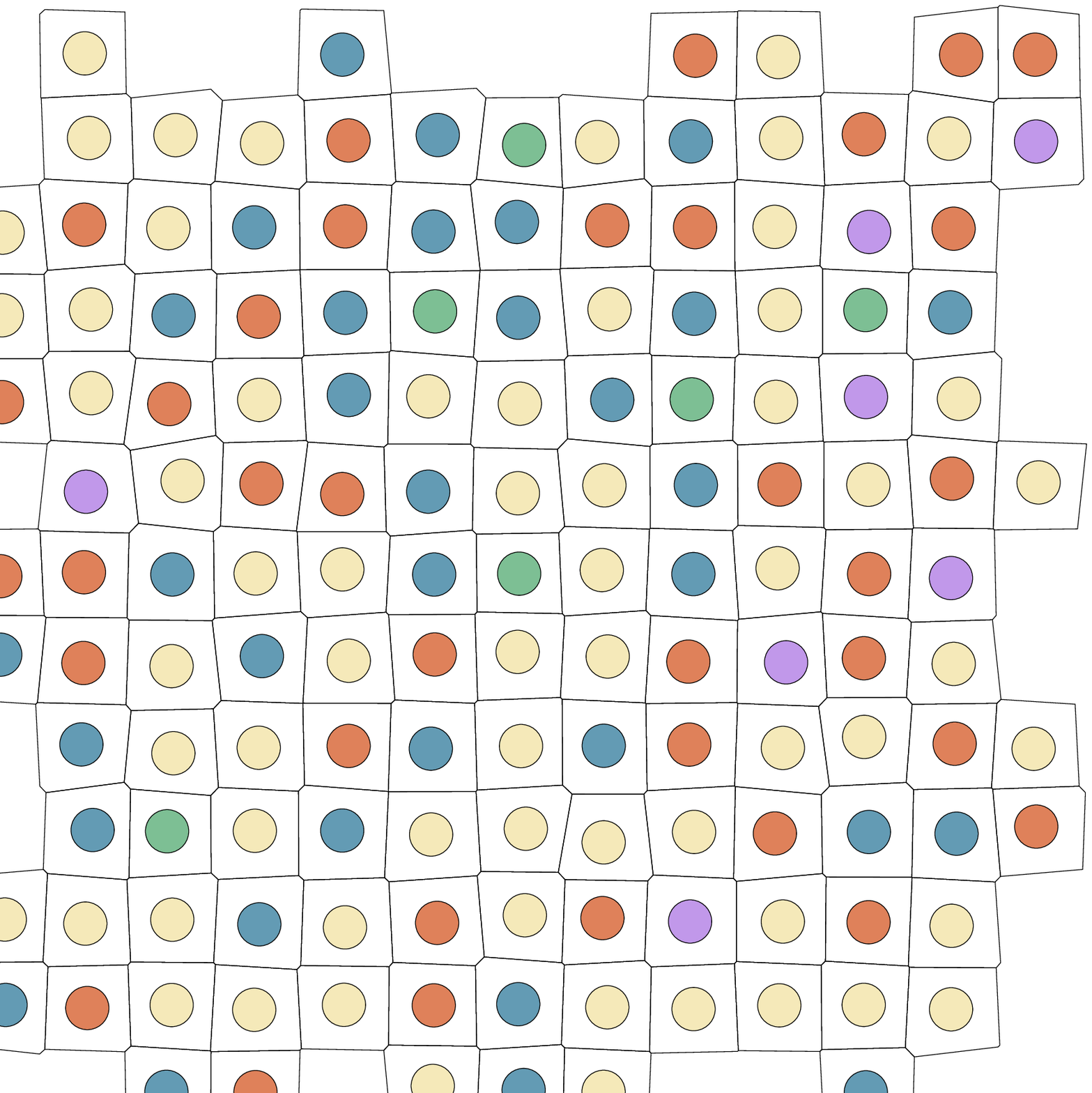}} \vspace{1mm}\\
(a) & (b) 
\end{tabular}$
\end{center}
\caption{Voronoi cells of (a) a square lattice and (b) a perturbed square lattice.  Atoms are colored according to the number of edges of their Voronoi cells: green, red, yellow, blue, and purple, for 4-8 edges, respectively.}
\label{perturbedlattice}
\end{figure}

We characterize the local ordering of individual atoms in atomic systems using their Voronoi cells.  Given a discrete set of points $S \subset \mathbb{R}^2$, their Voronoi tessellation is a partition of $\mathbb{R}^2$ into a countable number of sets called Voronoi cells \cite{1992okabe}.  The {\it Voronoi cell} of each point $s \in S$ is defined as
\begin{equation}
V(s) = \{x \in \mathbb{R}^2 \;|\; d(x, s) \leq d(x, s')\; \text{for all}\; s' \in S\},
\end{equation}
where $d$ is the standard Euclidean metric on $\mathbb{R}^2$.  Informally, the Voronoi cell of a point is the region of space closer to it than to any other point in $S$.  Figure \ref{perturbedlattice} shows Voronoi tessellations of unperturbed and perturbed square lattices.  

\setlength{\fboxsep}{0pt}

A point $x \in \mathbb{R}^2$ can be equidistant to multiple atoms and hence belong to multiple Voronoi cells.  A set of atoms is said to be in general position if no point in $\mathbb{R}^2$ belongs to more than three Voronoi cells or, equivalently, if no vertex of the Voronoi boundary graph is adjacent to more than three edges.  If atoms are not in general position, then some points in $\mathbb{R}^2$ belong to more than three Voronoi cells, and consequently some vertices in the Voronoi boundary graph are adjacent to four or more edges.  We call such vertices {\it unstable} because after perturbations, with probability one, they resolve into pairs of vertices connected by edges.  Every vertex in the Voronoi graph of the unperturbed square lattice is unstable, cf.~Fig.~\ref{perturbedlattice}(a).  After perturbations, with probability one, all vertices are stable and exactly three edges meet at each; cf.~Fig.~\ref{perturbedlattice}(b).  Throughout the paper, we use $V_{ij}$ to denote the Voronoi cell of the atom initially located at the lattice point $(i,j) \in \mathbb{Z}^2$; we define the random variable $X_{ij}$ to be the number of edges of $V_{ij}$ after perturbation.

\section{Voronoi chains}
\label{nchains}

We begin by considering Voronoi {\it n}-chains.  We define a {\bf Voronoi $n$-chain}, for $n \in \mathbb{N}$, to be a set of $n$ contiguous Voronoi cells in the same row or column.  To be concrete, throughout the paper we consider the Voronoi $n$-chain consisting of the Voronoi cells $\{ V_{0,0}, V_{1,0}, \ldots V_{n-1,0} \}$.  We thus consider the random variables $\{X_{0,0}, X_{1,0}, \ldots X_{n-1,0}\}$ and their joint distribution.  To simplify notation, we use $X_i$ to refer to $X_{i,0}$.  The probability measure on the space of perturbed lattices is stationary with respect to the group action $\mathbb{Z}^2$, as well as the group of rotations by $\frac{\pi}{2}$, and reflections, and so results are identical for any other set of contiguous Voronoi cells in a single row or column.  

In the small-perturbation limit, i.e.~$\sigma^2 \to 0$, the probability that a Voronoi cell will have fewer than 4 or more than 8 edges vanishes.  Therefore, in this limiting case, the combinatorics of a Voronoi $n$-chain can be described by a sequence of numbers $(X_0 = k_0, X_1 = k_1, \ldots, X_{n-1}=k_{n-1})$, where each $k_i \in \{4,5,6,7,8\}$ is the number of edges of a Voronoi cell.  We wish to calculate the probability that a Voronoi $n$-chain is described by a given sequence.  

We have shown previously \cite{leipold2016statistical} that the distribution of the random variable $X_{ij}$ for any $i,j$, i.e., the number of edges of the Voronoi cell of the atom initially located at $(i,j) \in \mathbb{Z}^2$, is given by:
\begin{equation}\label{P1}
P(X_{ij} = k)= 
\begin{cases} 
      [(\pi - \theta)/2\pi]^2 & k=4,8,\\
      \frac{\theta(\pi-\theta)}{\pi^2}, & k=5,7,\\
      \frac{\pi^2-2\pi\theta+3\theta^2}{2\pi^2},&  k=6,
   \end{cases}
\end{equation}
where $\theta = \arccos(-1/4)$.  We think of this distribution as that of Voronoi 1-chains.

\subsection*{2-chains} 
In considering the joint distribution $P(X_0=k_0, X_1=k_1)$, we note that $X_0$ and $X_1$, the number of edges of Voronoi cells $V_0$ and $V_1$, each depends on the manner in which four unstable vertices resolve.  In particular, before perturbations, each vertex in the Voronoi diagram belongs to four Voronoi cells.  After perturbations, with probability one, each unstable vertex resolves into a stable edge adjacent to a pair of atoms initially located $\sqrt{2}$ apart.  To facilitate discussion, we use $v_{ij}$ to refer to the unstable vertex initially located halfway between atoms $(i,j)$ and $(i-1,j-1)$.  We further use $v_{ij}^{\foos{}}$ to indicate the event that under perturbations the vertex $v_{ij}$ resolves into an edge belonging to Voronoi cells $V_{ij}$ and $V_{i-1,j-1}$, and $v_{ij}^{\hspace{0.2mm}\foo{}}$ to indicate the event that the newly-created edge belongs to Voronoi cells $V_{i-1,j}$ and $V_{i,j-1}$. For any unstable vertex $v_{ij}$, the events $v_{ij}^{\foo{}}$ and $v_{ij}^{\foos{}}$ are mutually exclusive and $P(v_{ij}^{\foo{}}) = P(v_{ij}^{\foos{}}) = \frac{1}{2}$, due to considerations of symmetry.

\begin{figure}
\begin{center}
\fbox{\begin{overpic}[trim={345mm 18mm 395mm 25mm},clip,width=0.65\linewidth]{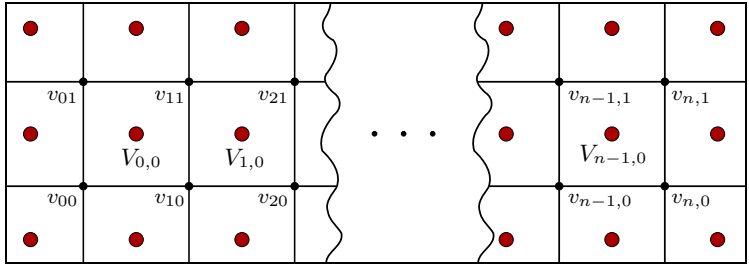}
\put (5.5,8) {\footnotesize $v_{00}$}
\put (5.5,22) {\footnotesize$v_{01}$}
\put (20,8) {\footnotesize$v_{10}$}
\put (20,22) {\footnotesize$v_{11}$}
\put (34,8) {\footnotesize$v_{20}$}
\put (34,22) {\footnotesize$v_{21}$}
\put (76,8) {\footnotesize$v_{n-1,0}$}
\put (76,22) {\footnotesize$v_{n-1,1}$}
\put (90,8) {\footnotesize$v_{n,0}$}
\put (90,22) {\footnotesize$v_{n,1}$}
\put (15.5,13.25) {\small $V_{0,0}$}
\put (29.5,13.25) {\small $V_{1,0}$}
\put (77.5,13.75) {\small $V_{n-1,0}$}
\end{overpic}}
\caption{Atoms and Voronoi cells in an unperturbed square lattice.  Before perturbations, the unstable vertices $v_{ij}$, indicated by small black circles, belong to Voronoi cells of four equidistant atoms.  After perturbations, each unstable vertex resolves in one of two ways, adding edges to two neighboring cells.}  
\label{unstablevertices}
\end{center}
\end{figure}

As each event $X_i = k_i$ depends on the manner in which four adjacent unstable vertices resolve, the event $\{X_0=k_0, X_1=k_1,\ldots, X_{n-1}=k_{n-1}\}$ depends on the manner in which $2(n+1)$ unstable vertices resolve (Fig.~\ref{unstablevertices}).  To understand the distribution of 2-chains, we thus need to consider the manner in which six unstable vertices resolve.  
\setlength{\fboxsep}{0pt}
Before considering the general case $P(X_0 =k_0, X_1=k_1)$, we look at the special case of $k_0=4$ and $k_1=8$.  

Although we previously calculated the unconditional probabilities $P(X_{ij}=4)$ and $P(X_{ij}=8)$, the events $X_0=4$ and $X_1=8$ are not independent, as both depend on the manner in which the vertices $v_{10}$ and $v_{11}$ resolve.  To determine $P(X_0 =4, X_1=8)$, we first calculate the unconditional probability $P(v_{00}^{\foo{}}, v_{01}^{\foos{}})$, and then the conditional probabilities $P(v_{10}^{\foos{}}, v_{11}^{\foo{}} |  v_{00}^{\foo{}}, v_{01}^{\foos{}})$ and $P(v_{20}^{\foo{}},v_{21}^{\foos{}} | v_{10}^{\foos{}}, v_{11}^{\foo{}})$.  Using the probabilistic chain rule, we thus have:
\begin{align}
\begin{split}
P(X_0 =4, X_1=8) &= P(v_{00}^{\foo{}}, v_{01}^{\foos{}}, v_{10}^{\foos{}}, v_{11}^{\foo{}}, v_{20}^{\foo{}},v_{21}^{\foos{}})\\
&= P(v_{00}^{\foo{}}, v_{01}^{\foos{}})  P(v_{10}^{\foos{}}, v_{11}^{\foo{}} |  v_{00}^{\foo{}}, v_{01}^{\foos{}})  P(v_{20}^{\foo{}},v_{21}^{\foos{}} | v_{10}^{\foos{}}, v_{11}^{\foo{}})\\
&= P(v_{00}^{\foo{}}, v_{01}^{\foos{}})  \frac{P(v_{00}^{\foo{}}, v_{01}^{\foos{}}, v_{10}^{\foos{}}, v_{11}^{\foo{}})}{P(v_{00}^{\foo{}}, v_{01}^{\foos{}})}  \frac{P(v_{10}^{\foos{}}, v_{11}^{\foo{}}, v_{20}^{\foo{}},v_{21}^{\foos{}})}{P(v_{10}^{\foos{}}, v_{11}^{\foo{}})} 
\end{split}
\label{eqn:4848}
\end{align}
The probabilities $P(v_{00}^{\foo{}}, v_{01}^{\foos{}}, v_{10}^{\foos{}}, v_{11}^{\foo{}})$ and $P(v_{10}^{\foos{}}, v_{11}^{\foo{}}, v_{20}^{\foo{}},v_{21}^{\foos{}})$ are exactly those of $P(X_0 = 4)$ and $P(X_1 = 8)$, respectively, given by Eq.~\ref{P1}.  As shown previously \cite{leipold2016statistical}, the events $v_{00}^{\foo{}}$ and $v_{01}^{\foos{}}$ are pairwise independent, as are $v_{10}^{\foos{}}$ and $v_{11}^{\foo{}}$, and so $P(v_{00}^{\foo{}}, v_{01}^{\foos{}}) = P(v_{00}^{\foo{}})P(v_{01}^{\foos{}}) = \frac{1}{4}$, and $P(v_{10}^{\foos{}}, v_{11}^{\foo{}})=P(v_{10}^{\foos{}})P( v_{11}^{\foo{}})=\frac{1}{4}$.  We therefore have:
\begin{equation}
P(X_0 =4, X_1=8) = 4 \left[ \frac{\pi-\theta}{2\pi}\right]^4 \approx 7.75 \times 10^{-3},
\label{eqn:48}
\end{equation}
where $\theta = \arccos(-1/4)$.  We also have $P(X_0 =8, X_1=4) = P(X_0 =4, X_1=8)$, given by symmetry.

Each of these two special cases $\{X_0 =4, X_1=8\}$ and $\{X_0 =8, X_1=4\}$ corresponds to exactly one choice of resolutions of the six adjacent vertices $\{v_{00}, v_{01}, v_{10}, v_{11}, v_{20},v_{21}\}$.  In general, however, multiple resolutions of the vertices $v_{ij}$ can result in identical values of $X_0$ and $X_1$.  As an example, the event $\{X_0 =5, X_1=5\}$ is a union of two disjoint events $\{v_{00}^{\foo{}}, v_{01}^{\foos{}}, v_{10}^{\foo{}}, v_{11}^{\foo{}}, v_{20}^{\foos{}},v_{21}^{\foo{}}\}$ and $\{v_{00}^{\foo{}}, v_{01}^{\foos{}}, v_{10}^{\foos{}}, v_{11}^{\foos{}}, v_{20}^{\foos{}},v_{21}^{\foo{}}\}$.  Calculations similar to those made above show that the probability of each one of these events is $4\left[\frac{\theta(\pi-\theta)}{4\pi^2} \right]^2$, resulting in 
\begin{align}
\begin{split}
P(X_0 =5, X_1=5) &= P(v_{00}^{\foo{}}, v_{01}^{\foos{}}, v_{10}^{\foo{}}, v_{11}^{\foo{}}, v_{20}^{\foos{}},v_{21}^{\foo{}}) + P(v_{00}^{\foo{}}, v_{01}^{\foos{}}, v_{10}^{\foos{}}, v_{11}^{\foos{}}, v_{20}^{\foos{}},v_{21}^{\foo{}})\\
&= 8\left[\frac{\theta(\pi-\theta)}{4\pi^2} \right]^2 \approx 2.97 \times 10^{-2}. 
\end{split}
\label{eqn:55}
\end{align}
Calculating $P(X_0 =k_0, X_1=k_1)$ for general $k_0, k_1$ requires determining the joint resolutions of the vertices $v_{ij}$ that result in those values of $X_i$, as well as their individual probabilities.  Table \ref{x1x2nums} summarizes the results of similar calculations for general $k_0$ and $k_1$.  

\begin{comment}
\begin{table}
\begin{center}
\mbox{\ensuremath{\begin{array}{|l|l|l|l|l|l|}
\hline
 k_{0}/k_{1}  &  4  &  5  &  6  &  7  &  8\\\hline
 4  &  0  &  0  &  \frac{1}{4}a^{2}b^{2}  &  \frac{1}{2}a^{3}b  &  \frac{1}{4}a^{4}\\\hline
 5  &  0  &  \frac{1}{2}a^{2}b^{2}  &  ab^{3}+\frac{1}{2}a^{3}b  &  \frac{3}{2}a^{2}b^{2}  &  \frac{1}{2}a^{3}b\\\hline
 6  &  \frac{1}{4}a^{2}b^{2}  &  ab^{3}+\frac{1}{2}a^{3}b  &  \frac{1}{2}a^{4}+a^{2}b^{2}+b^{4}  &  ab^{3}+\frac{1}{2}a^{3}b  &  \frac{1}{4}a^{2}b^{2}\\\hline
 7  &  \frac{1}{2}a^{3}b  &  \frac{3}{2}a^{2}b^{2}  &  ab^{3}+\frac{1}{2}a^{3}b  &  \frac{1}{2}a^{2}b^{2}  &  0\\\hline
 8  &  \frac{1}{4}a^{4}  &  \frac{1}{2}a^{3}b  &  \frac{1}{4}a^{2}b^{2}  &  0  &  0 \\\hline
\end{array}}}
\label{x1x2nums2}
\end{center}
\caption{The joint probability distribution $P(X_0 =k_0, X_1=k_1)$ of Voronoi 2-chains, with $a=\frac{\pi-\theta}{\pi}$, $b=\frac{\theta}{\pi}$, and $\theta = \arccos(-1/4)$.} 
\end{table}
\end{comment}

\def\arraystretch{1.08}
\begin{table}
\begin{center}
\mbox{\ensuremath{\begin{array}{|l|l|l|l|l|l|}
\hline
 k_{0}/k_{1}  &  4  &  5  &  6  &  7  &  8\\\hline
 4  &  0  &  0  &  4a^{2}b^{2}  &  8a^{3}b  &  4a^{4}\\\hline
 5  &  0  &  8a^{2}b^{2}  &  16ab^{3}+8a^{3}b  &  24a^{2}b^{2}  &  8a^{3}b\\\hline
 6  &  4a^{2}b^{2}  &  16ab^{3}+8a^{3}b  &  8a^{4}+16a^{2}b^{2}+16b^{4}  &  16ab^{3}+8a^{3}b  &  4a^{2}b^{2}\\\hline
 7  &  8a^{3}b  &  24a^{2}b^{2}  &  16ab^{3}+8a^{3}b  &  8a^{2}b^{2}  &  0\\\hline
 8  &  4a^{4}  &  8a^{3}b  &  4a^{2}b^{2}  &  0  &  0 \\\hline
\end{array}}}
\end{center}
\caption{The joint probability distribution $P(X_0 =k_0, X_1=k_1)$ of Voronoi 2-chains, with $a=\frac{\pi-\theta}{2\pi}$, $b=\frac{\theta}{2\pi}$, and $\theta = \arccos(-1/4)$.\label{x1x2nums}} 
\end{table}

\subsection*{General $n$-chains} 
\begin{figure}
\begin{center}
\begin{tabular}{cc}
(a) \fbox{\includegraphics[trim={15mm 15mm 15mm 15mm},clip,height=16mm]{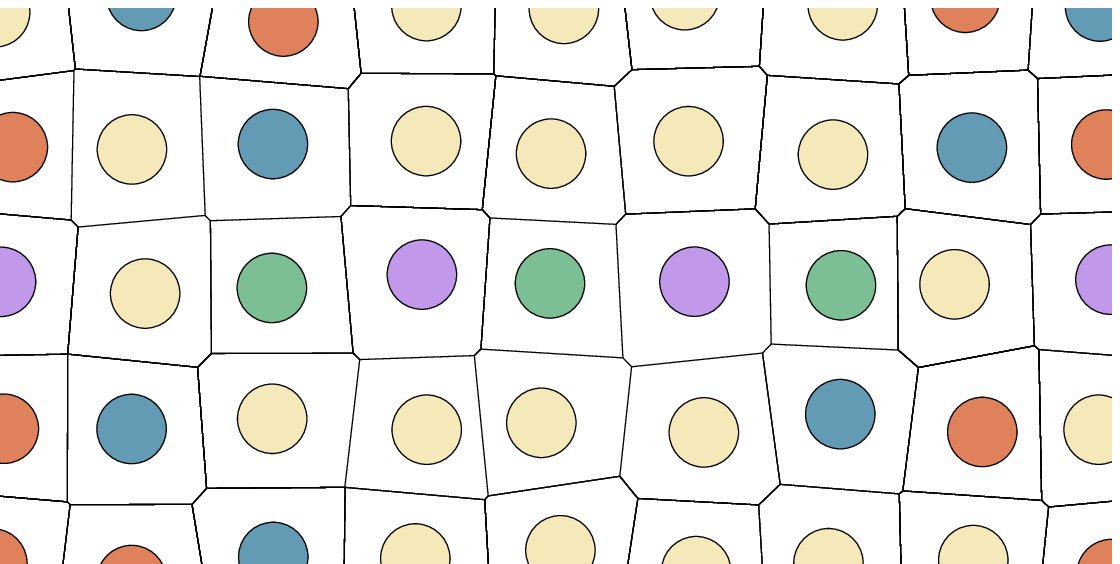}}&
(b) \fbox{\scalebox{-1}[1]{\includegraphics[trim={15mm 15mm 15mm 15mm},clip,height=16mm]{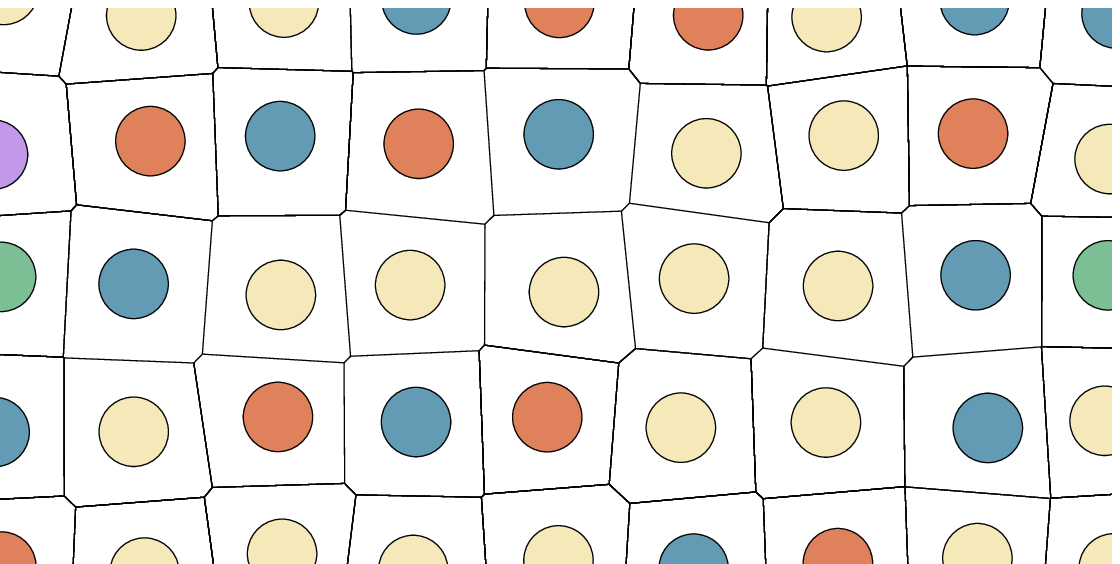}}}\vspace{2mm}\\
(c) \fbox{\includegraphics[trim={15mm 15mm 15mm 15mm},clip,height=16mm]{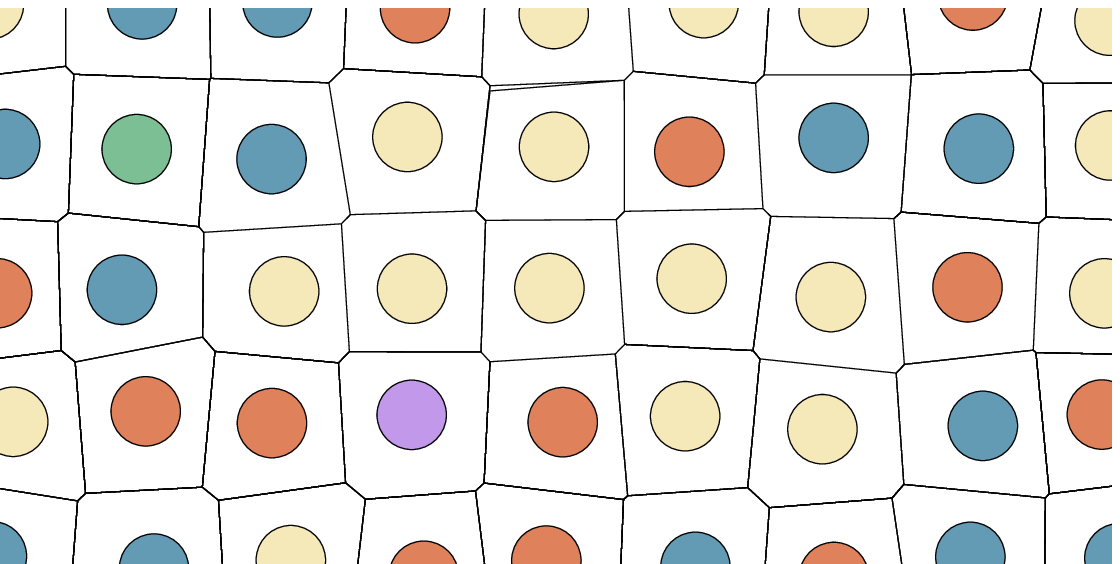}}&
(d) \fbox{\includegraphics[trim={15mm 15mm 15mm 15mm},clip,height=16mm]{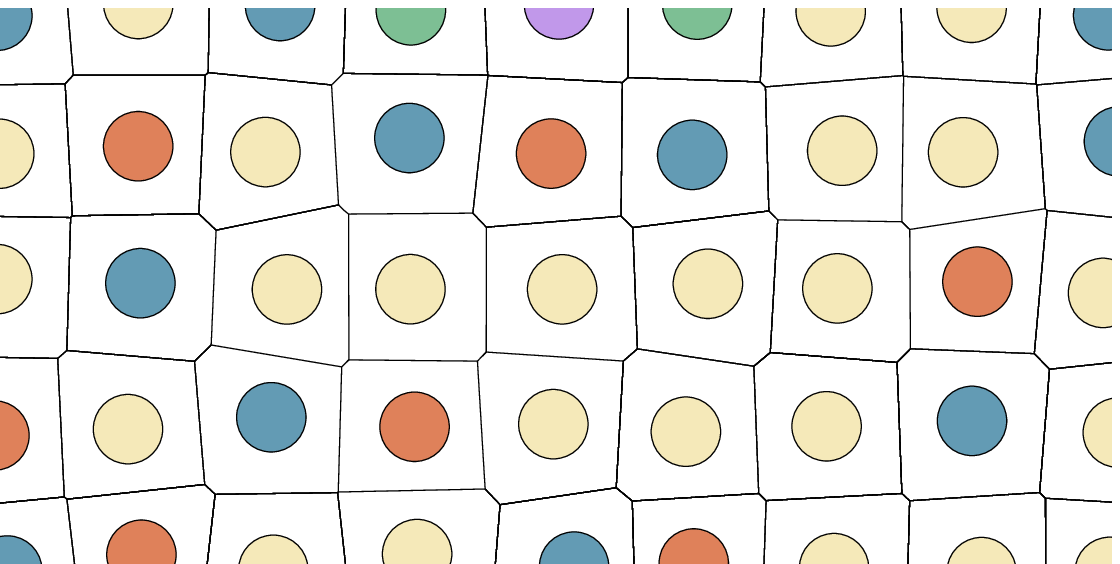}}
\end{tabular}
\caption{Voronoi chains consisting only of (a) quadrilateral and octagonal Voronoi cells, (b) hexagonal cells of the first kind, (c) and (d) hexagonal cells of the second kind, ordered differently.  Figures have been lightly edited to make the smaller edges more visible.}
\label{allChains}
\end{center}
\end{figure}

The calculations above immediately generalize to $n$-chains for all $n \in \mathbb{N}$.  We begin by extending our calculations above to $n$-chains consisting of 4- and 8-sided Voronoi cells.  In particular, we can generalize Eq.~\ref{eqn:4848} to:
\begin{align}
\begin{split}
P(X_0 =4, X_1=8, \ldots) 
&= P(v_{00}^{\foo{}}, v_{01}^{\foos{}}) 
\frac{P(v_{00}^{\foo{}}, v_{01}^{\foos{}}, v_{10}^{\foos{}}, v_{11}^{\foo{}})}{P(v_{00}^{\foo{}}, v_{01}^{\foos{}})} 
\frac{P(v_{10}^{\foos{}}, v_{11}^{\foo{}},v_{20}^{\foo{}}, v_{21}^{\foos{}})}{P(v_{10}^{\foos{}}, v_{11}^{\foo{}})} 
\cdots \\
&= P(v_{00}^{\foo{}})P(v_{01}^{\foos{}}) 
\frac{P(v_{00}^{\foo{}}, v_{01}^{\foos{}}, v_{10}^{\foos{}}, v_{11}^{\foo{}})}{P(v_{00}^{\foo{}})P(v_{01}^{\foos{}})} 
\frac{P(v_{10}^{\foos{}}, v_{11}^{\foo{}},v_{20}^{\foo{}}, v_{21}^{\foos{}})}{P(v_{10}^{\foos{}})P(v_{11}^{\foo{}})}
\cdots \\
&= \frac{1}{4} \left[ \frac{\pi-\theta}{\pi}\right]^{2n}.
\end{split}
\label{eqn48}
\end{align}
Due to combinatorial considerations, the events $\{X_i=4, X_{i+1}=4\}$ and $\{X_i=8, X_{i+1}=8\}$ are not possible for any $i \in \mathbb{Z}$, and so the Voronoi cells alternate in numbers of edges, with $X_i=4$ for $i$ even and $X_i=8$ for $i$ odd in this example.  The probability of a Voronoi $n$-chain consisting of only 4- and 8-sided Voronoi cells but beginning with an 8-sided Voronoi cell is identical, due to symmetry.  Figure \ref{allChains}(a) illustrates a Voronoi $n$-chain consisting of alternating 4- and 8-sided cells.

\vspace{3mm}

{\bf Hexagonal chains.} Voronoi $n$-chains consisting only of hexagonal cells can take one of several forms.  Before perturbations, each Voronoi cell has four edges, and four unstable vertices, each of which can resolve in a manner that adds one edge to the Voronoi cell.  Hexagonal cells result from perturbations in which either two adjacent or two opposite vertices resolve to add edges to the Voronoi cell.  We call those with opposite resolved edges hexagonal cells {\it of the first kind}, and those with adjacent resolved edges hexagonal cells {\it of the second kind}; numerous examples of both can be found in Fig.~\ref{perturbedlattice}.  Calculations similar to those presented in Ref.~\cite{leipold2016statistical} show that the probability of a Voronoi cell being hexagonal of the first kind is $2[(\pi - \theta)/2\pi]^2 \approx 8.80\%$, whereas that of it being hexagonal of the second kind is $[\theta/\pi]^2 \approx 33.69\%$.  Indeed, many more hexagonal cells of the second kind can be observed in Fig.~\ref{perturbedlattice} than those of the first kind.

{\it Hexagonal Voronoi chains of the first kind.} Hexagonal cells of each kind can be arranged to form Voronoi $n$-chains.  We consider three such types of chains.  The Voronoi chain illustrated in Fig.~\ref{allChains}(b) consists entirely of hexagonal Voronoi cells of the first kind.  Figures \ref{allChains}(c) and (d) show Voronoi chains consisting entirely of hexagonal Voronoi cells of the second kind, arranged in two combinatorially distinct patterns.  
\begin{comment}
\begin{figure}
\begin{center}
\begin{tabular}{ccc}
\fbox{\includegraphics[trim={36mm 23mm 36mm 23mm},clip,height=12mm]{SQ-500-05-29.eps}}&
\fbox{\includegraphics[trim={36mm 23mm 36mm 23mm},clip,height=12mm]{SQ-500-05-13.eps}}&
\fbox{\includegraphics[trim={36mm 23mm 36mm 23mm},clip,height=12mm]{SQ-500-05-17.eps}}\\
(a)&(b)&(c)
\end{tabular}
\caption{(a) Voronoi chain consisting only of hexagonal cells of the first kind; (b) and (c) chains consisting of hexagonal cells of the second kind, ordered differently.}
\label{chain666}
\end{center}
\end{figure}
\end{comment}
To compute the probability that $n$ consecutive Voronoi cells will be hexagonal cells of the first kind, we consider the special case illustrated in Fig.~\ref{allChains}(b).  This chain is defined as the event $\{v_{ij}^{\foos{}} : 0 \leq i \leq n, 0 \leq j \leq 1\}$.  We again use the probabilistic chain rule:
\begin{align}
\begin{split}
P(v_{ij}^{\foos{}} : 0 \leq i \leq n, 0 \leq j \leq 1) &= P(v_{00}^{\foos{}}, v_{01}^{\foos{}})  P(v_{10}^{\foos{}}, v_{11}^{\foos{}} |  v_{00}^{\foos{}}, v_{01}^{\foos{}}) \cdot \cdot \cdot  P(v_{n,0}^{\foos{}},v_{n,1}^{\foos{}} | v_{n-1,0}^{\foos{}}, v_{n-1,1}^{\foos{}})\\
&= \frac{1}{4} \left[ \frac{\pi-\theta}{\pi}\right]^{2n}.
\end{split}
\label{eqn666a}
\end{align}
The probability of the event $\{v_{ij}^{\foo{}} : 0 \leq i \leq n, 0 \leq j \leq 1\}$ is identical, by symmetry, and so the total probability of a Voronoi $n$-chain consisting entirely of hexagonal cells of the first kind is $\frac{1}{2} \left[ \frac{\pi-\theta}{\pi}\right]^{2n}$.  

{\it Hexagonal Voronoi chains of the second kind.} Voronoi $n$-chains consisting only of hexagonal cells of the second kind can appear in two forms.  The one illustrated in Fig.~\ref{allChains}(c) corresponds to the event $\{v_{ij}^{\foos{}}, i \textrm{ even}; v_{ij}^{\foo{}}, i \textrm{ odd}\}$, whereas that illustrated in Fig.~\ref{allChains}(d) corresponds to the event $\{v_{ij}^{\foo{}}, j \textrm{ even}; v_{ij}^{\foos{}}, j \textrm{ odd}\}$.  Informally, either entire rows or entire columns of $v_{ij}$ resolve in alternating directions.  Calculations similar to those above show that the probability of each of these two kinds of chains are identical and equal to 
\begin{equation}
\frac{1}{4} \left[ \frac{\theta}{\pi} \right]^{2n}.
\label{eq:hex2}
\end{equation}
Two additional events, $\{v_{ij}^{\foo{}}, i \textrm{ even}; v_{ij}^{\foos{}}, i \textrm{ odd}\}$ and $\{v_{ij}^{\foos{}}, j \textrm{ even}; v_{ij}^{\foo{}}, j \textrm{ odd}\}$, have the same probabilities.  Hence, the total probability of a Voronoi $n$-chain consisting entirely of hexagonal cells of the second kind is $\left[ \frac{\theta}{\pi} \right]^{2n}$.

{\it General hexagonal chains.}  General hexagonal Voronoi chains consist of a mixture of Voronoi cells of the first and second kind, as can be observed in Fig.~\ref{perturbedlattice}.  To obtain the probability of the event $\{X_0=6, \ldots, X_{n-1}=6\}$, we consider two possible sub-events.  In particular, we observe that for every $i$, a pair of vertices $v_{i,0}$ and $v_{i,1}$ can either resolve in parallel (e.g., $v_{i,0}^{\foos{}}, v_{i,1}^{\foos{}}$) or in orthogonal (e.g., $v_{i,0}^{\foos{}}, v_{i,1}^{\foo{}}$) directions.  If they resolve in parallel, then the joint resolution of the vertices adds one edge to each of the Voronoi cells on either side of them.  If they resolve in orthogonal directions, then the joint resolution of the vertices adds two edges to the Voronoi cell on one side, and none to that on the other.  If a pair of orthogonally-resolving vertices were adjacent to a pair of vertices resolving in parallel, then the Voronoi cell in between would necessarily have either 5 or 7 edges. Hence, general hexagonal Voronoi chains must either consist entirely of pairs of vertices that resolve in parallel, or else consist entirely of pairs of vertices that resolve orthogonally.  Moreover, if an hexagonal Voronoi chain consists of orthogonally resolving pairs of vertices, then all pairs must resolve in the same manner --- if $v_{0,0}^{\foos{}}, v_{0,1}^{\foo{}}$, then $v_{i,0}^{\foos{}}, v_{i,1}^{\foo{}}$ for all $0 \leq i \leq n$.  Otherwise, some Voronoi cells would have either 4 or 8 edges.  

To facilitate discussion, we introduce the notation $v^{\foop{}}_i = \{v_{i,0}^{\foos{}}, v_{i,1}^{\foos{}} \} \cup \{v_{i,0}^{\foo{}}, v_{i,1}^{\foo{}} \}$ to denote the event that vertices $v_{i,0}$ and $v_{i,1}$ resolve in parallel, and $v^{\foog{}}_i = \{v_{i,0}^{\foos{}}, v_{i,1}^{\foo{}} \} \cup \{v_{i,0}^{\foo{}}, v_{i,1}^{\foos{}} \}$ to denote the event that the vertices resolve orthogonally.  We can thus rewrite the event of general hexagonal Voronoi chains as a disjoint union $\{X_0=6, \ldots, X_{n-1}=6\} = \{v^{\foop{}}_i, 0 \leq i \leq n\} \cup \{v^{\foog{}}_i, 0 \leq i \leq n\}$.

Hexagonal Voronoi chains solely consisting of pairs of vertices that resolve in parallel include the Voronoi chain illustrated in Figs.~\ref{allChains}(b) and (c), and mixtures of the two.  To compute the probability that a Voronoi chain will consist only of pairs of vertices that resolve in parallel, we again employ the probabilistic chain rule.  The initial pair of vertices can resolve in parallel in one of two manners, each with probability $P(v_{0,0}^{\foos{}}, v_{0,1}^{\foos{}}) = P(v_{0,0}^{\foo{}}, v_{0,1}^{\foo{}}) = \frac{1}{4}$.  
The conditional probability that vertices $v_{i+1,0}$ and $v_{i+1,1}$ resolve in parallel, given that $v_{i,0}$ and $v_{i,1}$ resolve in parallel is then given by:
\begin{equation}
P(v^{\foop{}}_{i+1} | v^{\foop{}}_i) = P(v_{i+1,0}^{\foos{}}, v_{i+1,1}^{\foos{}} | v^{\foop{}}_i) + P(v_{i+1,0}^{\foo{}}, v_{i+1,1}^{\foo{}}  | v^{\foop{}}_i).
\end{equation}
Due to symmetry of the problem, it is sufficient to consider the case where vertices $v_{i+1,0}$ and $v_{i+1,1}$ resolve in parallel in one particular manner.  We thus have:
\begin{align}
\begin{split}
P(v^{\foop{}}_{i+1} | v^{\foop{}}_i) &= P(v_{i+1,0}^{\foos{}}, v_{i+1,1}^{\foos{}} | v_{i,0}^{\foos{}}, v_{i,1}^{\foos{}}) + P(v_{i+1,0}^{\foo{}}, v_{i+1,1}^{\foo{}}  | v_{i,0}^{\foos{}}, v_{i,1}^{\foos{}})\\
 &= \left[\frac{\pi-\theta}{\pi} \right]^2 + \left[\frac{\theta}{\pi} \right]^2,
\end{split}
\end{align}
where the conditional probabilities are determined using methods described earlier.  Using the probabilistic chain rule, we conclude that the probability that all $n+1$ pairs of vertices resolve in parallel ways is given by the product $P(v^{\foop{}}_0)=\frac{1}{2}$ with the $n$ conditional probabilities that subsequent pairs of vertices also resolve in parallel:
\begin{equation}
P(v^{\foop{}}_i, 0 \leq i \leq n) = \frac{1}{2}\left(\left[\frac{\pi-\theta}{\pi} \right]^2 + \left[\frac{\theta}{\pi} \right]^2 \right)^n
\label{eqn:parallel}
\end{equation}

A hexagonal Voronoi chain solely consisting of pairs of vertices that resolve orthogonally is illustrated in Fig.~\ref{allChains}(d).  The probability of this event has been calculated previously; cf.~Eq.~\ref{eq:hex2}.  In particular, we have
\begin{align}
\begin{split}
P(v^{\foog{}}_i, 0 \leq i \leq n) &= P(v_{ij}^{\foo{}}, j \textrm{ even}; v_{ij}^{\foos{}}, j \textrm{ odd})+P(v_{ij}^{\foos{}}, j \textrm{ even}; v_{ij}^{\foo{}}, j \textrm{ odd})\\
&=\frac{1}{2}\left[ \frac{\theta}{\pi} \right]^{2n}.
\end{split}
\label{eqn10}
\end{align}

The probability that all Voronoi cells are hexagons is thus given by the sum of Eqs.~\ref{eqn:parallel} and \ref{eqn10}:
\begin{equation}
P(X_0=6, \ldots, X_{n-1}=6) = \frac{1}{2}\left[ \frac{\theta}{\pi} \right]^{2n} + \frac{1}{2}\left(\left[\frac{\pi-\theta}{\pi} \right]^2 + \left[\frac{\theta}{\pi} \right]^2 \right)^n.
\end{equation}

\begin{comment}
{\bf Chains with 5's and 7's.}  As a final example of patterned Voronoi chains, we consider those chains consisting only of pentagons and heptagons.  Similar calculations as those above show that:
\begin{equation*}
P(X_0=5, X_1=7, \ldots, X_{n-1}) = 
\begin{cases}
4\cdot 2^{5(n-1)/2} \left( \frac{\pi-\theta}{4\pi^2} \right)^{n}&\text{$n$ odd}\\
3\sqrt{2}\cdot 2^{5(n-1)/2} \left( \frac{\pi-\theta}{4\pi^2} \right)^{n}&\text{$n$ even;}
\end{cases}
\end{equation*}
Figure \ref{chain57} shows a chain consisting only of pentagons and heptagons.
\begin{figure}
\begin{center}
\fbox{\includegraphics[trim={30mm 16mm 30mm 16mm},clip,height=12mm]{SQ-500-05-31.eps}}
\caption{A Voronoi chain consisting only of pentagonal (red) and heptagonal (blue) cells.}
\label{chain57}
\end{center}
\end{figure}
\end{comment}

{\bf Arbitrary $n$-chains.} Calculating probabilities of general $n$-chains is complicated by the many resolutions of the vertices $v_{ij}$ that result in particular sequences $(k_0, k_1,\ldots k_{n-1})$.  It is not clear whether anything interesting can be stated about the general case.  In theory, however, each probability can be computed exactly by consideration of the resolutions resulting in a particular sequence $(k_i)$, and summing over their probabilities.

\section{Voronoi blocks}

\begin{figure}[b]
\begin{center}
\begin{tabular}{ccc}
\fbox{\includegraphics[trim={15mm 16mm 15mm 14mm},clip,width=0.28\linewidth]{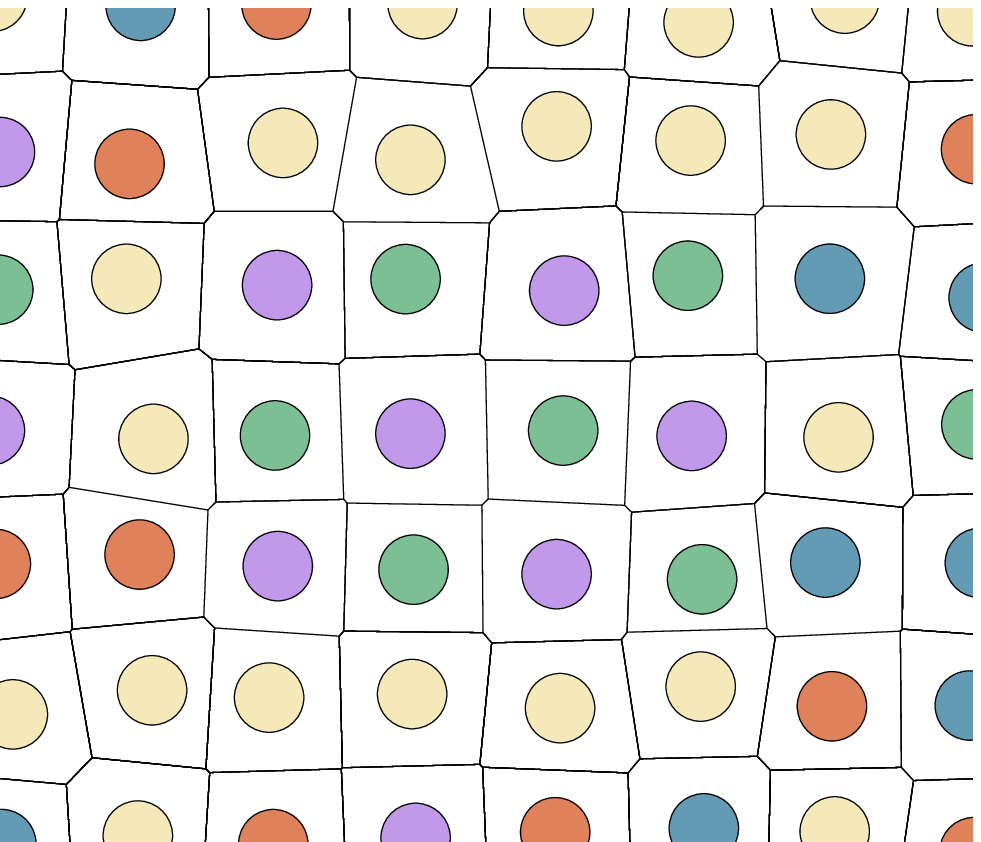}}&
\fbox{\includegraphics[trim={15mm 15mm 15mm 15mm},clip,width=0.28\linewidth]{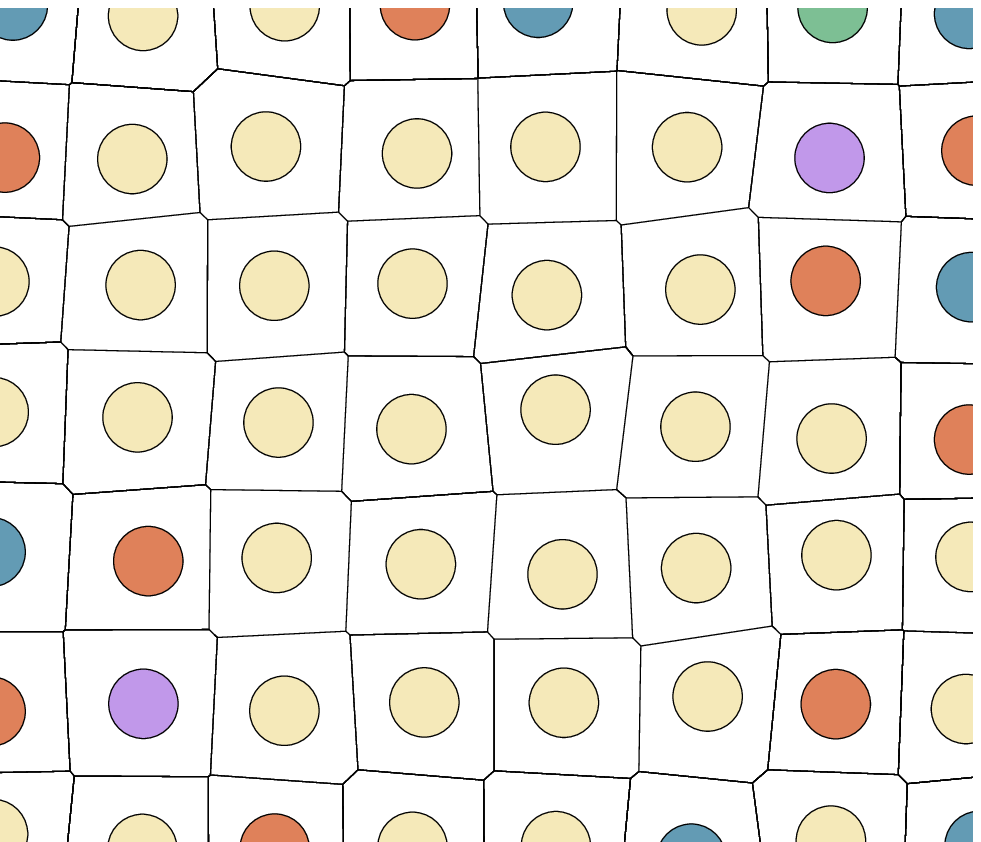}}&
\fbox{\includegraphics[trim={15mm 15mm 15mm 15mm},clip,width=0.28\linewidth]{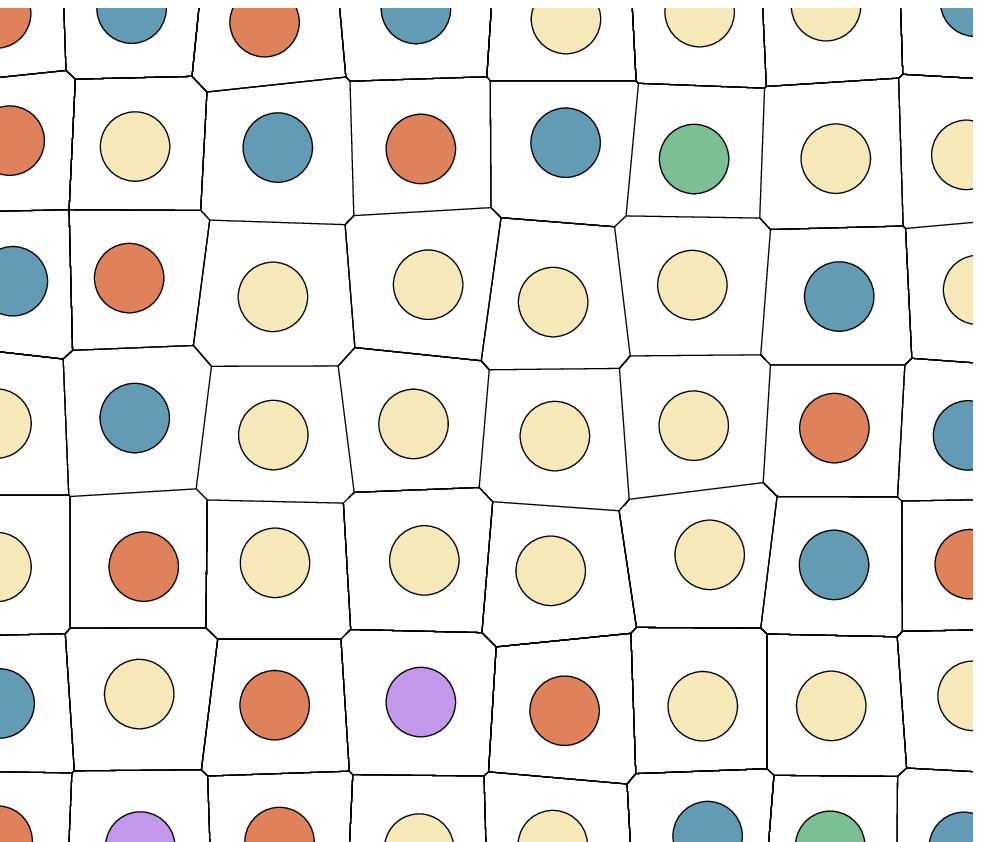}}\\
(a)&(b)&(c)
\end{tabular}
\caption{Voronoi blocks consisting only of (a) quadrilateral and octagonal cells, (b) hexagonal cells of the first kind, and (c) hexagonal cells of the second kind.  Figures have been lightly edited to make the smaller edges more visible.}
\label{blocks}
\end{center}
\end{figure}

The method used here to calculate the distribution of Voronoi chains can be extended to two-dimensional arrays, which we call {\it blocks}, of Voronoi cells.  In particular, we consider $m \times n$ arrays of Voronoi cells with given combinatorial features.  These patterns of Voronoi cells can be considered as regions of alternate order inside an ambient square lattice phase, resulting from small random perturbations.  We consider three examples: Voronoi blocks consisting only of quadrilateral and octagonal cells, those consisting only of hexagonal cells of the first kind, and those consisting only of hexagonal cells of the second kind.  Examples of these Voronoi blocks are illustrated in Fig.~\ref{blocks}.
As in the case of $n$-chains, for the sake of concreteness we consider the particular block consisting of the Voronoi cells $\{ V_{i,j} : 0 \leq i < n, 0 \leq j < m\}$.  As mentioned previously, the probability measure on the space of perturbed lattices is stationary with respect to symmetries of $\mathbb{Z}^2$, and so results are identical for all Voronoi blocks.  

Our general strategy is to treat the $m$ rows as individual Voronoi $n$-chains, the unconditional probabilities of which have been computed in Section \ref{nchains}.  We then calculate the probability that row $j+1$ will be a Voronoi $n$-chain of a given type, conditional on row $j$ being a Voronoi $n$-chain of that type.  Since these conditional probabilities are identical for all $m-1$ rows, the probability that all $m$ rows are Voronoi $n$-chains of a given type is given by the product of these probabilities with the unconditional probability that the first row is a Voronoi chain of the given type.  

{\bf Hexagonal of first kind.} Voronoi blocks consisting only of hexagonal cells of the first kind can be characterized by the joint event $\{ v_{ij}^{\foos{}} : 0 \leq i \leq n, 0 \leq j \leq m\}$, or else $\{ v_{ij}^{\foo{}}\}$ for the same vertices; cf.~Fig.~\ref{blocks}(b).  We have previously computed the probability of an hexagonal $n$-chain of the first kind using Eq.~\ref{eqn666a}: $P(v_{ij}^{\foos{}} : 0 \leq i \leq n, 0 \leq j \leq 1) = \frac{1}{4} \left[ \frac{\pi-\theta}{\pi}\right]^{2n}$.  Such chains are formed by two adjacent rows of vertices, all of which resolve in the same way. This result can be used to calculate the conditional probability that a second row of vertices all resolve in a particular direction, given that a first row of vertices all resolve in that direction.  Formally,
\begin{eqnarray}
P(v_{01}^{\foos{}}, v_{11}^{\foos{}}, \ldots, v_{n,1}^{\foos{}}|v_{00}^{\foos{}}, v_{10}^{\foos{}}, \ldots, v_{n,0}^{\foos{}}) &=& \frac{P(v_{ij}^{\foos{}} : 0 \leq i \leq n, 0 \leq j \leq 1)}{P(v_{00}^{\foos{}}, v_{10}^{\foos{}}, \ldots, v_{n,0}^{\foos{}})}.
\end{eqnarray}
Since the resolution of all unstable vertices in a single row or column are independent, the probability in the denominator is $P(v_{00}^{\foos{}}, v_{10}^{\foos{}}, \ldots, v_{n,0}^{\foos{}}) = P(v_{00}^{\foos{}})P(v_{10}^{\foos{}}) \ldots P(v_{n,0}^{\foos{}}) = \left(\frac{1}{2}\right)^{n+1}$, giving us
\begin{eqnarray}
P(v_{01}^{\foos{}}, v_{11}^{\foos{}}, \ldots, v_{n,1}^{\foos{}}|v_{00}^{\foos{}}, v_{10}^{\foos{}}, \ldots, v_{n,0}^{\foos{}}) &=& 2^{n-1} \left[ \frac{\pi-\theta}{\pi}\right]^{2n}.
\end{eqnarray}
Since rows of hexagonal cells of the first kind are formed by consecutive rows all of whose vertices resolve in the same direction, this conditional probability of the vertices can be interpreted as the conditional probability that row $j+1$ consists of hexagonal cells of the first kind, conditioned on row $j$ consisting of Voronoi cells of the same kind.  

The probability that an $m\times n$ Voronoi block consists entirely of hexagonal cells of the first kind is then given by the product of the unconditional probability of the first row consisting of such cells, given by Eq.~\ref{eqn666a}, and the conditional probabilities of the $m-1$ additional rows resolving in an identical manner.  This reduces to:
\begin{equation}
P(v_{ij}^{\foos{}} : 0 \leq i \leq n, 0 \leq j \leq m) = 2^{(m-1)(n-1)-2}\left[\frac{\pi-\theta}{\pi}\right]^{2mn}.
\label{mnblock6a}
\end{equation}
Due to considerations of symmetry, the probability of the event $\{ v_{ij}^{\foo{}} : 0 \leq i \leq n, 0 \leq j \leq m \}$ is identical, and so the total probability that an $m\times n$ Voronoi block consists entirely of hexagonal cells of the first kind is twice this number.  
\vspace{5mm}

{\bf Hexagonal of second kind.}  Similar calculations can be made to obtain the probability that a Voronoi block consists entirely of hexagonal cells of the second kind.  Although Voronoi chains consisting entirely of hexagonal cells of the second kind can appear in one of two distinct patterns, when these chains are stacked together to form blocks, they appear in only one pattern, isomorphic under symmetries of $\mathbb{Z}^2$.  To simplify analysis, we begin by considering the case where the first row of Voronoi cells is described by the event $\{v_{ij}^{\foos{}}, i \textrm{ even}; v_{ij}^{\foo{}}, i \textrm{ odd}\}$, illustrated in Fig.~\ref{blocks}(c).  Analysis of each of the three similar events $\{v_{ij}^{\foo{}}, i \textrm{ even}; v_{ij}^{\foos{}}, i \textrm{ odd}\}$, $\{v_{ij}^{\foos{}}, j \textrm{ even}; v_{ij}^{\foo{}}, j \textrm{ odd}\}$, and $\{v_{ij}^{\foo{}}, j \textrm{ even}; v_{ij}^{\foos{}}, j \textrm{ odd}\}$ is identical.

We have already seen in Eq.~\ref{eq:hex2} that the probability of the event $\{v_{ij}^{\foos{}}, i \textrm{ even}; v_{ij}^{\foo{}}, i \textrm{ odd}\}$ is $\frac{1}{4} \left[ \frac{\theta}{\pi} \right]^{2n}$.  This result can be used to calculate the conditional probability that a row of Voronoi cells will be an $n$-chain consisting entirely of hexagonal cells of the second kind, given that the row before it was an $n$-chain consisting entirely of hexagonal cells of the second kind:
\begin{equation}
P(v_{i,j+1}^{\foos{}}, i \textrm{ even}; v_{i,j+1}^{\foo{}}, i \textrm{ odd} | v_{ij}^{\foos{}}, i \textrm{ even}; v_{ij}^{\foo{}}, i \textrm{ odd} ) = 2^{n-1}\left[\frac{\theta}{\pi} \right]^{2n}.
\end{equation}
The probability that an $m\times n$ Voronoi block consists entirely of hexagonal cells of the second kind is then the product of the probability of the first row being a Voronoi $n$-chain of this kind multiplied by the $m-1$ conditional probabilities of the remaining rows also being Voronoi $n$-chains of this kind:
\begin{equation}
P(v_{ij}^{\foo{}}, i \textrm{ odd}; v_{ij}^{\foos{}}, i \textrm{ even})  = 2^{(m-1)(n-1)-2}\left[\frac{\theta}{\pi}\right]^{2mn}.
\end{equation}
Since there are four distinct manners in which the first row can be hexagonal of the second kind, each with the same probability, the total probability of an $m\times n$ block consisting entirely of hexagonal cells of the second kind is four times this number.  
\vspace{5mm}

{\bf Quadrilateral-octagonal phase.}  We last consider Voronoi blocks consisting entirely of quadrilaterals and octagons.  The probability that the first row will consist of alternating quadrilaterals and octagons is given by Eq.~\ref{eqn48}.  The conditional probability that row $j+1$ will be a Voronoi $n$-chain consisting of alternating octagons and quadrilaterals, conditioned on row $j$ being constituted of these Voronoi cells, is:
\begin{align}
\begin{split}
P(X_{0,j+1} =8, X_{1,j+1} =4, \ldots | X_{0,j} =4, X_{1,j} =8, \ldots) &= \frac{P(X_0 =4, X_1 =8, \ldots)}{P(v_{ij}^{\foo{}}, i \textrm{ even}; v_{ij}^{\foos{}}, i \textrm{ odd})}\\
&= 2^{n-1}\left[ \frac{\pi-\theta}{\pi} \right]^{2n}.
\end{split}
\end{align}
The probability that an $m\times n$ Voronoi block consists entirely of alternating quadrilateral and octagonal Voronoi cells is then given by the product of the probability of the first row having this structure, with the conditional probabilities of the remaining $m-1$ rows having this structure:
\begin{equation}
P(X_{ij} = 4, i+j \textrm{ even}; X_{ij} = 8, i+j \textrm{ odd}) = 2^{(m-1)(n-1)-2}\left[\frac{\pi-\theta}{\pi}\right]^{2mn}.
\label{bulk48}
\end{equation}
Due to considerations of symmetry, the probability $P(X_{ij} = 8, i+j \textrm{ even}; X_{ij} = 4, i+j \textrm{ odd})$ is identical.  Therefore, the probability that an $m \times n$ block of Voronoi cells will consist of only quadrilateral and octagonal cells is twice this number.  We note that the probability given by Eq.~\ref{bulk48} for the quadrilateral-octagonal phase is identical to that given in Eq.~\ref{mnblock6a} for Voronoi blocks of hexagonal cells of the first kind.

These three examples of bulk ``phases'' are the only possible simple repeating patterns in perturbed square lattices.  In particular, it is not possible to have blocks, or even chains, consisting only of quadrilateral Voronoi cells, or only octagonal Voronoi cells.  It is similarly not possible to have three consecutive pentagons or 7-gons.  While it is possible to have more complex repeating patterns, those analyzed here appear to be the only ones with small repeating unit.

\section{Voronoi clusters}
\label{vclusters}

We finally consider contiguous sets of Voronoi cells of certain types that do not necessarily lie in rectangular blocks, but instead belong to generally-shaped connected regions.  We define two Voronoi cells to be {\it neighbors} if they share a boundary edge in common.  A {\bf Voronoi $n$-cluster} is then a connected set of $n$ Voronoi neighbors.  In this section we consider Voronoi $n$-clusters consisting only of hexagonal cells of the first kind, only of hexagonal cells of the second kind, and only of quadrilateral and octagonal cells.  How does the average number of such $n$-clusters per atom depend on $n$?  In contrast to Voronoi chains and blocks, exact results for Voronoi clusters are difficult to obtain due to their irregular shapes and due to the correlation between neighboring Voronoi cells.  We, therefore, employ numerical simulations to study them.  

\begin{figure}
\begin{center}
\begin{tabular}{ccc}
\fbox{\includegraphics[trim={28mm 28mm 28mm 28mm},clip,width=0.3\linewidth]{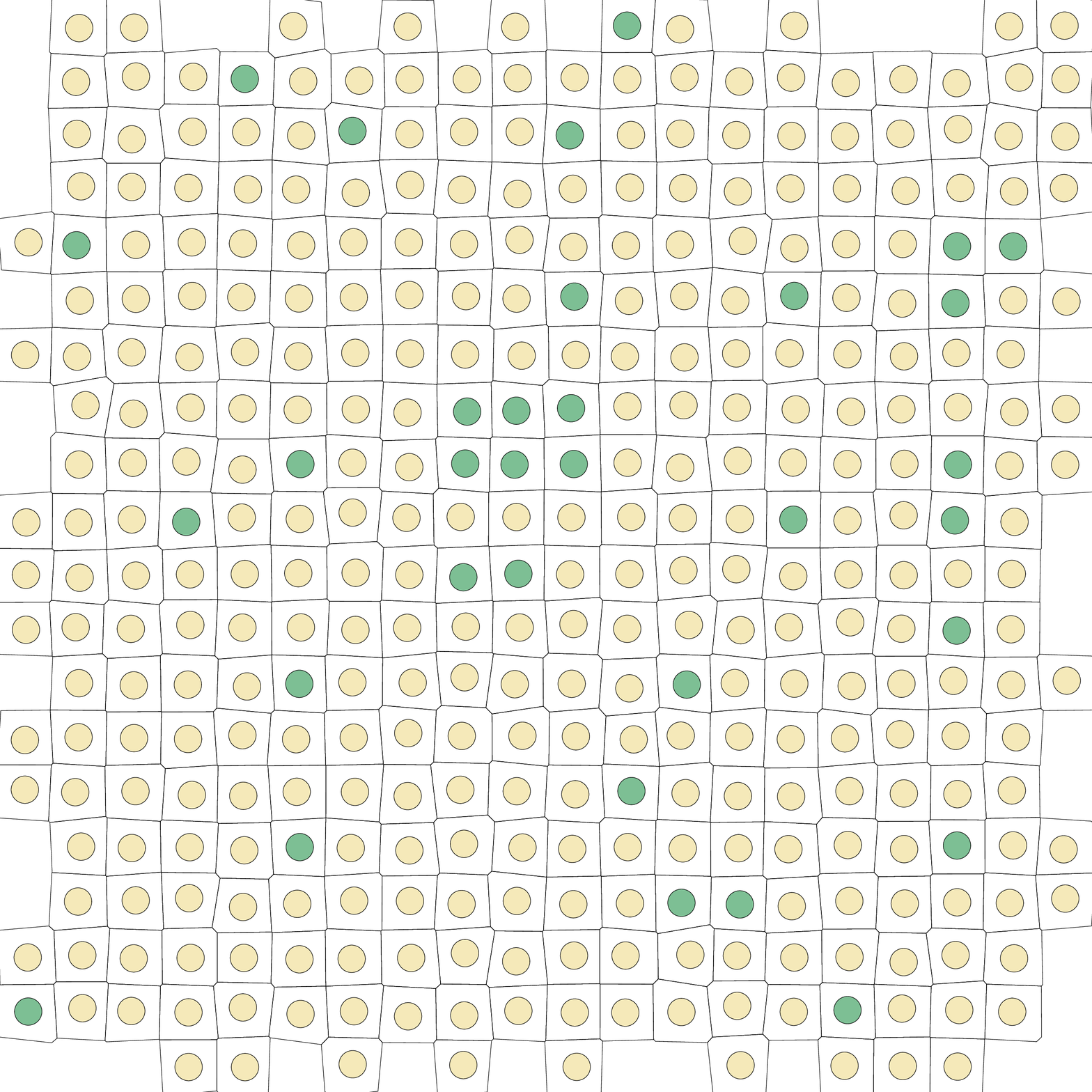}}&
\fbox{\includegraphics[trim={28mm 28mm 28mm 28mm},clip,width=0.3\linewidth]{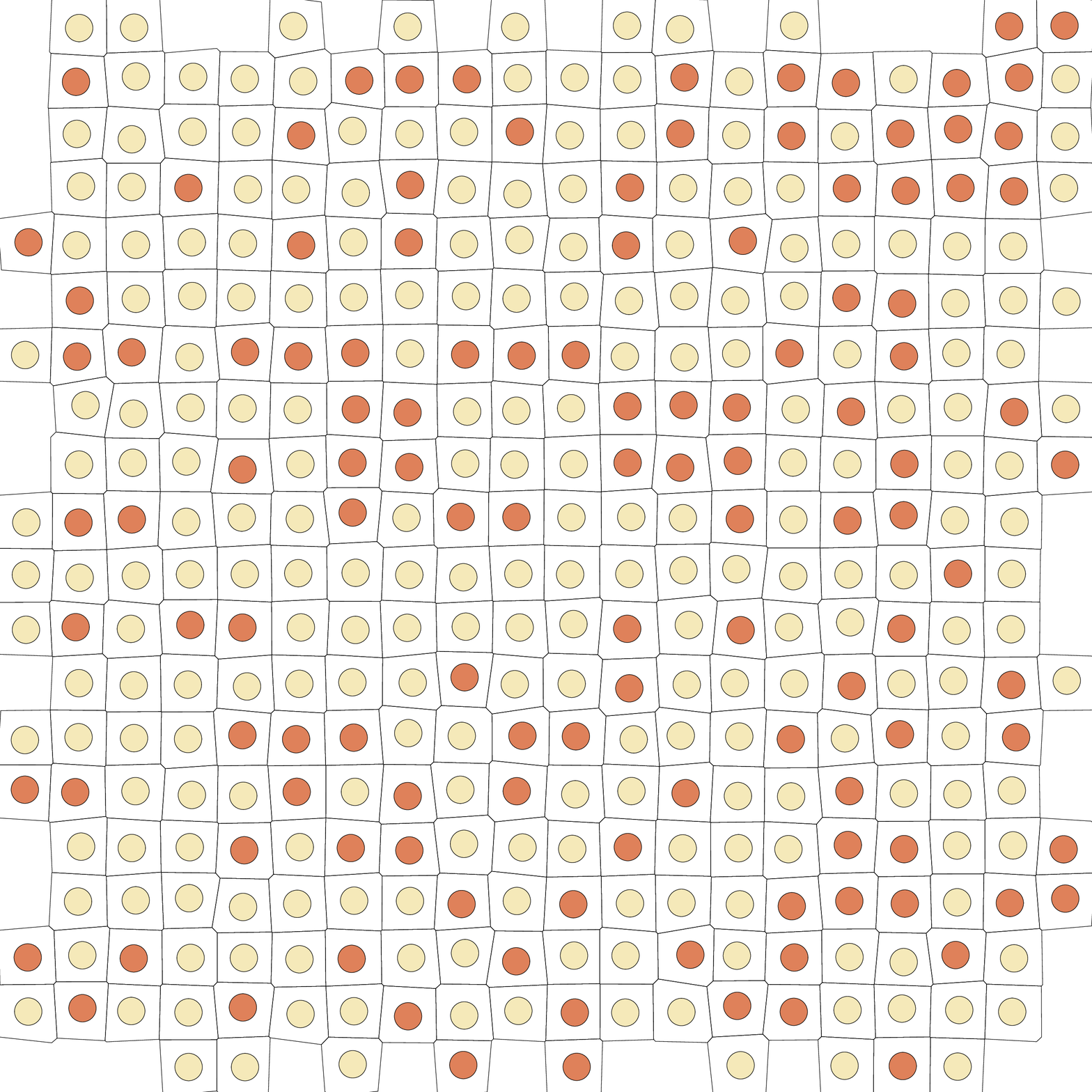}}&
\fbox{\includegraphics[trim={28mm 28mm 28mm 28mm},clip,width=0.3\linewidth]{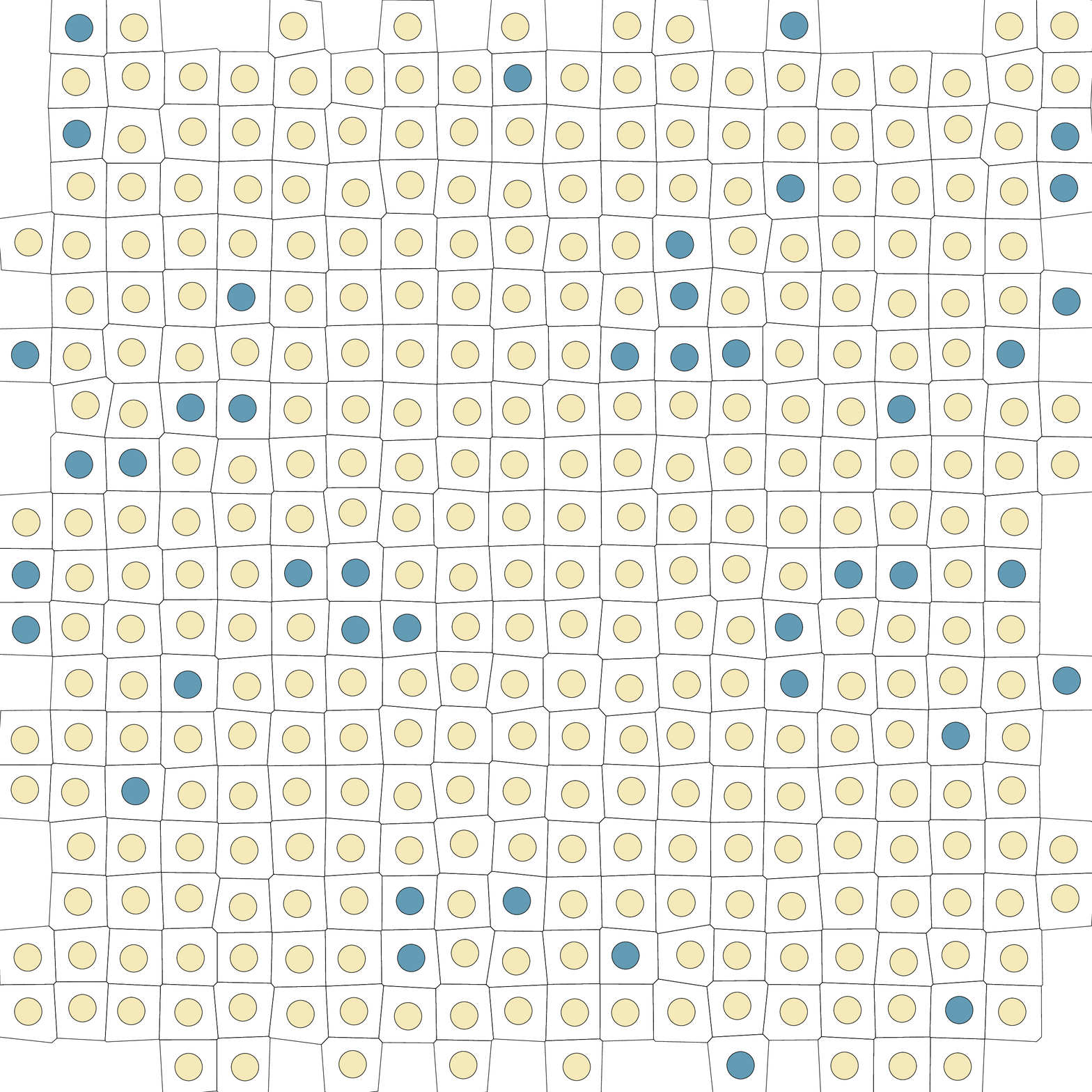}}\\
(a)&(b)&(c)
\end{tabular}
\caption{Perturbed lattices with highlighted Voronoi clusters consisting only of (a) hexagonal cells of the first kind, (b) hexagonal cells of the second kind, and (c) quadrilateral and octagonal cells.}
\label{fig:data:clusters}
\end{center}
\end{figure}

{\it Simulation details.} We began with a $1000 \times 1000$ grid of points with integer coordinates and periodic boundary conditions, and perturbed the $x$ and $y$ coordinates of each point by independently sampling from a normal distribution with mean zero and standard deviation $\sigma = 0.05$.  We then used the {\it VoroTop} software \cite{vorotop} to calculate the distribution of cluster sizes in these systems; some additional programming was necessary to distinguish the different kinds of hexagonal cells.  This experiment was repeated 500 times, for a total of 500 million sample Voronoi cells in the combined systems.  For each system, we computed the distribution of Voronoi clusters consisting only of hexagonal cells of the first kind, clusters consisting only of hexagonal cells of the second kind, and clusters consisting only of quadrilateral and octagonal cells.  Figure \ref{fig:data:clusters} illustrates a single perturbed lattice, and separately highlights each of these three kinds of clusters.

\begin{figure}
\begin{center}
\begin{overpic}[height=0.4\linewidth]{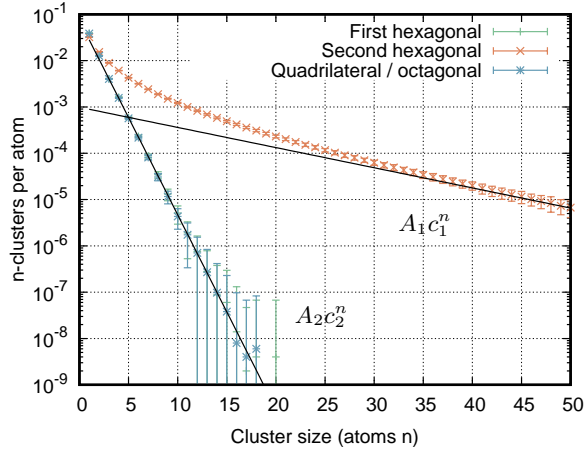}
\put (66,36) {\small $A_1c_1^n$}  
\put (49,21) {\small$A_2c_2^n$}
\end{overpic}
\caption{Average number of Voronoi $n$-clusters per atom as a function of $n$ for various kinds of structures; error bars show standard error from the mean.  Fitted curves of the form $Ac^n$ describe the asymptotic behavior for large $n$, with $A_1=9.85\times 10^{-4}$, $c_1=9.05\times 10^{-1}$, $A_2=7.38\times 10^{-2}$, and $c_2=3.81\times 10^{-1}$.  Estimated values of $A$ and $c$ were obtained by a least-squares fit for data for large $n$ in the samples.}
\label{hexclusters}
\end{center}
\end{figure}

{\it Results.} Figure \ref{hexclusters} plots the average number of Voronoi $n$-clusters per atom as a function of $n$. The numbers of $n$-clusters consisting of quadrilaterals and octagons appear identical to those of hexagonal Voronoi cells of the first kind.  This might not be surprising, given that the probabilities of Voronoi blocks consisting entirely of one or the other are identical, as calculated earlier.  Voronoi $n$-clusters of hexagonal cells of the second kind are considerably more common, consistent with results regarding Voronoi chains and blocks; this is particularly noticeable for large $n$.  The number of $n$-clusters per atom appears to asymptotically approach $Ac^n$ as $n\rightarrow \infty$, for constants $A$ and $c$.  The two fitted curves illustrated in Fig.~\ref{hexclusters} are both of this form.  

\section{Discussion and conclusions}

This paper considers two-dimensional perturbed square lattices --- simple models of low-temperature crystals --- and analyzes the distribution of subregions with prescribed combinatorial features.  Voronoi topology analysis of atomic systems has grown in importance for material characterization, both in simulation and in experiment \cite{2020morley}.  Although this paper focuses on the two-dimensional square lattice, similar methods can be applied to unstable lattices in three dimensions.  Indeed a similar general approach has been used in recent work \cite{lazar2015topological} to characterize defects and disorder in three-dimensional atomic systems. 

The results presented here take no account of the particular details that govern the evolution of physical systems.  It might prove worthwhile to explore how more realistic systems evolving under interatomic potentials and described by equilibrium statistical mechanics compare with that studied here.  In particular, would such systems have fewer, or more, regions of alternative ordering than appear here?  Would shapes of the clusters appear more compact instead of sprawling?  

Furthermore, the results here are exact for the low-temperature model in which perturbations are small relative to the lattice constant.  Exploring the distribution of hexagonal clusters in high-temperature crystals might provide valuable insight into non-equilibrium phase transitions.  For example, although the square lattice is commonly mechanically unstable, as noted earlier, it can be stabilized under certain thermodynamic conditions \cite{lozovik2019spontaneous,marcotte2011optimized}.  When those conditions are changed, these systems may undergo a square-lattice to hexagonal phase transition \cite{damasceno2009pressure}.   How does a hexagonal phase nucleate within a square lattice one?  The results here can be interpreted to show that two different kinds of hexagonal ordering arise in localized regions even below the critical transition point.  Voronoi blocks of hexagonal Voronoi cells of either the first or second kind, both of which can be found with non-vanishing probability in square-lattice systems, are ideal candidates for nucleation of this new phase.  As hexagonal Voronoi blocks of the second kind are substantially more common, at least in the non-physical system considered here, we might reasonably expect that nucleation occurs preferentially at these locations.  Alternatively, it is possible that under interatomic potentials of interest, hexagonal Voronoi blocks of the first kind are more common.  Understanding the structure of perturbed square lattices can thus shed light on both equilibrium and non-equilibrium questions concerning the nature of phase transitions.

\section*{Acknowledgments} 
EAL acknowledges the generous support of the United States -- Israel Binational Science Foundation (BSF), Jerusalem, Israel through grant number 2018170.

\appendix

\section{Percolation}

We briefly consider a classic percolation problem \cite{kirkpatrick1973percolation,stauffer2018introduction} on the perturbed square lattice.  Each atom is colored with probability $p$, and we consider for what values of $p$ does there exist a colored percolating cluster.  The site-percolation threshold is estimated to be $0.593$ in the unperturbed square lattice \cite{derrida1985corrections}, which can be understood as the case in which $\sigma^2 =0 $.  In that case, every Voronoi cell has exactly four neighbors.  In the Poisson-Voronoi case, which might be thought of as the limit in which $\sigma^2 \rightarrow \infty$, Bollob{\'a}s and Riordan showed that the critical percolation threshold is exactly one half \cite{bollobas2006critical}.  The perturbed lattice considered here is the limit when $\sigma^2 \rightarrow 0$.

We used numerical simulations to estimate $p_c$ for the perturbed square lattice.  Since a sharp percolation threshold is typically only well-defined in the infinite-system limit, we simulated systems of increasingly larger sizes.  We began with an $n \times n$ square grid of points with integer coordinates and periodic boundary conditions, using $n \in \{100, 200, 400, 800\}$.  In each sample, we perturbed the $x$ and $y$ coordinates of each point by independently sampling from a normal distribution with mean zero and standard deviation $\sigma = 0.001$.  For values of $0.4 \leq p < 0.6$, we independently colored each Voronoi cell with probability $p$.  To determine whether a single cluster percolated, we identified the largest cluster in the system and determined whether for every $0 \leq i < n$, there was at least one atom with initial $x$-coordinate $i$ belonging to that cluster, and similarly for $y$-coordinates for every $0 \leq j < n$.  We repeated this experiment 1000 times for each $p$ and for each linear dimension $n$, recording the fraction of samples in which the largest colored cluster percolated.

\begin{figure}[h]
\begin{center}
\includegraphics[height=0.4\linewidth]{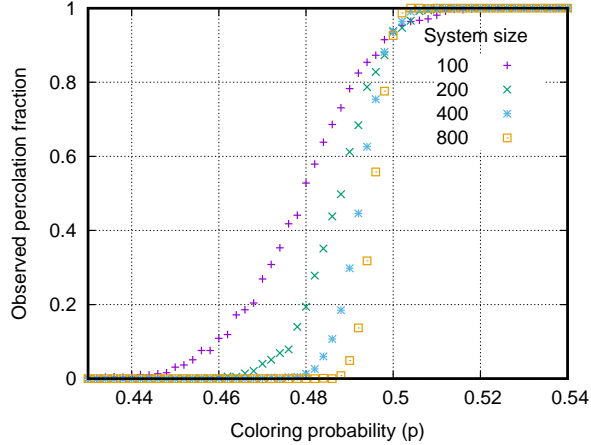}\\
\caption{Estimates of percolation for various system sizes $n \times n$.  As the system size increases, the probability of observing a percolating cluster appears to approach a step function at $p=\frac{1}{2}$.}
\label{filler}
\end{center}
\end{figure}

Figure \ref{filler} reports the observed finite-system probability of percolation given a probability $p$ of independently coloring each Voronoi cell, for different system sizes.  As the system sizes increase, the probability of percolation as a function of $p$ appears to converge to a step function, with the critical transition threshold located at $p_c = \frac{1}{2}$.  This threshold is strictly less than that of the unperturbed lattice, since the additional edges formed at unstable vertices function to connect otherwise unconnected regions.  Further analysis is necessary to compute this value exactly.

\bibliographystyle{ieeetr}

\end{document}